\documentclass[journal,twoside,web]{ieeecolor}
\usepackage{generic}
\usepackage{cite}
\usepackage{amsmath,amssymb,amsfonts}
\usepackage{algorithmic}
\usepackage{graphicx}
\usepackage{algorithm,algorithmic}
\usepackage{hyperref}
\hypersetup{hidelinks}
\usepackage{textcomp}
\def\BibTeX{{\rm B\kern-.05em{\sc i\kern-.025em b}\kern-.08em
    T\kern-.1667em\lower.7ex\hbox{E}\kern-.125emX}}
\usepackage{mathrsfs}
\usepackage{epstopdf}
\usepackage{tikz}
\usetikzlibrary{calc}
\usetikzlibrary{matrix, positioning, fit, calc}
\usepackage{tkz-euclide}
\usepackage{epsfig} 
\usepackage{enumerate}
\usepackage{multicol}
\usepackage{pst-plot,pst-eucl}
\usepackage{booktabs}

\newtheorem{assumption}{Assumption}

\newtheorem{remark}{Remark}

\newtheorem{lemma}{Lemma}

\newtheorem{theorem}{Theorem}

\let\oldforall\forall
\renewcommand{\forall}{\oldforall \, }
\let\oldexist\exists
\renewcommand{\exists}{\oldexist \: }

\newcommand{\G}{\mathscr{G}}
\newcommand{\V}{\mathcal{V}}

\newcommand{\M}{\mathcal{M}}
\newcommand{\E}{\mathscr{E}}
\newcommand{\F}{\mathscr{F}}
\newcommand{\K}{\mathcal{K}}
\newcommand{\I}{\mathcal{I}}

\newcommand{\N}{\mathcal{N}}
\newcommand{\U}{\mathcal{U}}
\newcommand{\Q}{\mathcal{Q}}

\newcommand{\J}{\mathcal{J}}
\renewcommand{\O}{\mathcal{O}}

\newcommand{\LL}{\mathbf{L}}
\newcommand{\DD}{\mathbf{D}}
\renewcommand{\L}{\mathcal{L}}

\usepackage{accents}

\newcommand{\real}{\mathbb{R}}

\newcommand{\naturals}{\mathbb{N}}

\newcommand{\lb}{[\![}
\newcommand{\rb}{]\!]}
\newcommand{\sys}[2]{\lb #1, #2\rb}
\newcommand{\innerp}[2]{\left \langle #1, #2\right \rangle}

\definecolor{gg}{rgb}{0.25,0.25,0.25}
\newcommand\mc[1]{{\color{black} #1}}

\definecolor{rg}{rgb}{0.3,0.7,0}

\title{Partitioned robustness analysis of networks with uncertain links
\thanks{Supported in part by the 
Australian Research Council (DP210103272), and 
National Science and Technology Council, Taiwan (113-2918-I-110-003, 113-2221-E-110-048-MY3).}}
\author{Simone Mariano, Chung-Yao Kao, and Michael Cantoni
\thanks{%
S.~Mariano is with the Institut d'ing\'{e}nierie et de management, Grenoble INP--UGA, 38031 Grenoble Cedex 1, France. E-mail: {\tt simomariano95@gmail.com} 
\\ 
\indent C.-Y.~Kao is with the Department of Electrical Engineering, National Sun Yat-sen University, Kaohsiung 80424, Taiwan. E-mail: {\tt cykao@ mail.nsysu.edu.tw} 
\\ 
\indent M.~Cantoni (corresponding author) is with the Department of Electrical and Electronic Engineering, The University of Melbourne, Parkville 3010, Australia. E-mail: {\tt cantoni@unimelb.edu.au}}}

\begin{document}


\maketitle
\begin{abstract}
Network \mc{robustness to} link \mc{perturbations is studied from an input-output perspective}. The main result is a set of integral quadratic constraints (IQCs) that imply robust stability of the uncertain network dynamics. The model dependency of each IQC is localized \mc{according to an overlapping} edge-partition \mc{for} the network graph. \mc{The choice of partition over the admissible set} affords flexibility \mc{in balancing} conservativeness of the distributed certificate \mc{against its verification complexity. This is} illustrated by a numerical example \mc{for a network with sampled-data links}.
\end{abstract}

\section{Introduction}

\mc{With relevance in resource (power/water/data) distribution, robot co-ordination, biology, etc.,} the study of \mc{network dynamics} has a long history in \mc{the} control~\mc{community}~\cite{siljak1978large,moylan1978stability,vidyasagar1981input,arcak2016networks}. 

An input-output \mc{perspective is adopted} here. The aim is to progress scalable analysis of \mc{network robustness to link uncertainty}. \mc{Following the preliminary work}~\cite{MCLinks,MCNodes}, the \mc{uncertainty is viewed as perturbing the ideal links known to yield nominal network stability, which is common in the study of cyber-physical systems~\cite{heemels2010networked,thomas2021frequency}.} \mc{Similar to~\cite{langbort2004distributed,jonsson2010scalable,andersen2014robust}, integral quadratic constraints (IQCs) are used to encapsulate the link uncertainty, and analysis of the associated network feedback interconnection hinges} on the IQC stability theorem~\cite{megretski1997system}. Model structure \mc{is leveraged} to decompose the resultant monolithic robustness 
certificate in a way related to~\cite{lestas2010,khong2014scalable,pates2016scalable}. \mc{However,} this existing work does not apply directly \mc{because the loop transformation associated with the perturbation model of link uncertainty obscures network structure, as elaborated in Section~\ref{sec:ideal+IQC}; see Remark~\ref{rem:unstruct}.}

The main contribution stems from a new structured coprime factorization of the \mc{nominal network} model with ideal links. This \mc{enables} reformulation of the monolithic certificate \mc{and decomposition into a sufficient collection of reduced IQCs.} Each \mc{IQC} only depends on localized model parameters \mc{in accordance with an overlapping} edge-partition \mc{for} the network graph. \mc{The choice of partition over the admissible set allows for} a trade-off \mc{between the} conservativeness and \mc{the verification complexity of the correspondingly} distributed certificate. \mc{The admissible set contains the finest possible partition from~\cite{MCNodes}, and coarser ones that yield reduced conservativeness with increased computation. The associated flexibility constitutes the  advance over~\cite{MCNodes}, which supersedes~\cite{MCLinks}. A summary appears in~\cite{CanKao2026}, without proofs, and with narrower scope.}

Preliminaries are presented next. The \mc{network} model and monolithic robustness \mc{certificate} from~\cite{MCNodes} \mc{are extended to accommodate unstable agents and links in Section~\ref{sec:ideal+IQC}}. The coprime factor \mc{based} decomposition, \mc{and semi-definite programming based verification complexity,} are given in Section~\ref{sec:mainresults} \mc{for admissible edge-partitions}. \mc{To conclude, the robustness of a virtual mass-spring-damper network to aperiodic sampling-induced link-delay is considered in Section~\ref{sec:numsec}. Significant verification complexity reduction is observed for a small penalty in conservativeness with coarse partitions. Further partition refinements yield diminished computational benefit, and degraded conservativeness of the robustness certificate.} 

\section{Preliminaries} 

\subsection{Basic notation} \label{subsec:notation}
 Let $[i:j]$ denote $\{ k\in\mathbb{N}~|~i\leq k\leq j\}$ with natural order; this is abbreviated $[j]$ when $i=1\leq j$; it is empty when $j=0$. $\mathbb{R}^{r\times q}$ consists of real matrices with $r\in\mathbb{N}$ rows and $q\in\mathbb{N}$ columns; 
 $\mathbb{R}^{r}:=\mathbb{R}^{r\times 1}$. The identity is denoted
 $I_r\in\mathbb{R}^{r\times r}$. All entries of $\mathbf{0}_{r,q}\in\mathbb{R}^{r\times q}$, and~$\mathbf{1}_{r,q}\in\mathbb{R}^{r\times q}$, are respectively zero, and~one. 
Given $X\in \mathbb{R}^{r\times q}$, $i\in[r]$, and $j\in[q]$, column $j$ is $X_{(\cdot,j)} \in \mathbb{R}^{r}$,
row $i$ is $X_{(i,\cdot)}\in\mathbb{R}^{1\times q}$, and the scalar in row $i$, column $j$, is $X_{(i,j)}\in\mathbb{R}$. Given ordered set $\I=\{i_1,\ldots,i_{|\I|}\}$ of cardinality $|\I|$, and $X_i\in\mathbb{R}^{r_i\times q_i}$ for $i\in\I$, block diagonal $\bigoplus_{i\in\I} X_i=X_{i_1}\oplus\cdots\oplus X_{i_{|\I|}} \in \mathbb{R}^{r\times q}$, where $r=\sum_{i\in\I} r_i$, $q=\sum_{i\in\I} q_i$. The transpose of
$X \in\mathbb{R}^{r\times q}$ is $X^\prime\in\mathbb{R}^{q\times r}$, and $Y^{-1}\in\mathbb{R}^{r\times r}$ is the inverse of non-singular $Y\in\mathbb{R}^{r\times r}$. Given $W=W^\prime,Z=Z^\prime\in\mathbb{R}^{r\times r}$, the respective expressions $Z \!\succeq\! W$ and $Z\!\succ\! W$, mean $(\forall v\in\mathbb{R}^r) ~v^\prime (Z-W) v \geq 0$ and $(\exists \epsilon\in\mathbb{R}_{>0})(\forall v\in\mathbb{R}^r)~ v^\prime (Z-W) v \geq \epsilon v^\prime v$.

\subsection{Signals and systems}
$\LL_{2\,}^r$ is the Hilbert space of signals $v:\mathbb{T} \rightarrow \real^r$ with finite energy $\|v\| := \langle v, v \rangle^{1/2}$, where $\langle v, u \rangle := \int_{\mathbb{T}} v(t)^\prime u(t) \, \mathrm{d}t$ and $\mathbb{T}:=\mathbb{R}_{\geq 0}$. The extended space $\LL_{2e}^r\supset \LL_2^r$ comprises  $v:\mathbb{T}\rightarrow\real^r$ such that $\boldsymbol{\pi}_{\tau}(v)\in\LL_{2\,}^r$ for all $\tau\in\mathbb{T}$, where $(\boldsymbol{\pi}_\tau(v))(t):=f(t)$ for $t\in\{t\in\mathbb{T}:t<\tau\}$, and $0$ elsewhere.

Given $G_i:\LL_{2e}^{q_i}\rightarrow \LL_{2e}^{r_i}$, $i\in\I$, and the isometric isomorphisms $\LL_{2e}^{q} \simeq \LL_{2e}^{q_1} \times \cdots \times \LL_{2e}^{q_{\smash{|\I|}}}$ and $\LL_{2e}^{r} \simeq \LL_{2e}^{r_1} \times \cdots \times \LL_{2e}^{r_{\smash{|\I|}}}$, where $q=\sum_{i\in\I} q_i$ and $r=\sum_{i\in\I} r_i$,
the direct sum
$\bigoplus_{i\in\I} G_i := ((u_i)_{i\in\I} \mapsto (G_i(u_i))_{i\in\I}): \LL_{2e}^{q}\rightarrow \LL_{2e}^{r}$. The composition
$G_i\circ G_j:=(u\mapsto G_i(G_j(u))):\LL_{2e}^{q_j}\rightarrow\LL_{2e}^{r_i}$; for linear maps, the $\circ$ and argument brackets are dropped. The map $G:\LL_{2e}^q\rightarrow \LL_{2e}^r$ is a {\em system} if it is causal, in the sense that $(\forall \tau\in\mathbb{T})~\boldsymbol{\pi}_\tau(G(u)) = \boldsymbol{\pi}_\tau(G (\boldsymbol{\pi}_\tau(u)))$, and $G(0)=0$. If $(\exists c\in\real_{>0})~(\forall u\in\LL_2^q)~ \|G(u)\| \leq c \|u\|$ as well, then the system is {\em stable}. For convenience, pointwise multiplication by a matrix is denoted by the matrix. Such linear maps are causal and bounded; i.e., stable. When (possibly non-causal) linear $G:\LL_{2e}^q\rightarrow \LL_{2e}^r$ has bounded gain, $G^*:\LL_{2e}^r\rightarrow \LL_{2e}^q$ denotes the Hilbert adjoint; i.e., $\langle y, Gu \rangle = \langle G^* y, u \rangle$.

The feedback interconnection of system $G:\LL_{2e}^q\rightarrow \LL_{2e}^r$ and system $\varDelta:\LL_{2e}^r \rightarrow \LL_{2e}^q$ is {\em well-posed} if 
for all 
$(d_y,d_u)\in\LL_{2e}^r\times\LL_{2e}^q$, there exists unique 
$(y,u)\in\LL_{2e}^r\times\LL_{2e}^q$ such that 
\begin{align} \label{eq:feedback1}
y = G(u) + d_y, \qquad u = \varDelta(y) + d_u, 
\end{align}
and $[\![G,\varDelta]\!] := ((d_y,d_u) \mapsto (y,u))$ is causal. In this case, $(I_r-G\circ\varDelta)$ and $(I_q - \varDelta\circ G)$ are bijective with causal inverses denoted by $(I_r-G\circ\varDelta)^{-1}$ and $(I_q - \varDelta\circ G)^{-1}$.  
\begin{theorem}[IQC Robust Stability~\cite{megretski1997system}]
\label{thm:robust_stability}
Given the stable system $\varDelta:\LL_{2e}^r \rightarrow \LL_{2e}^q$ and bounded self-adjoint multiplier $\varPi:\LL_{2\,}^r\times\LL_{2\,}^q \rightarrow \LL_{2\,}^r\times \LL_{2\,}^q$ (i.e., $\varPi=\varPi^*$), suppose  
\begin{equation*}
(\forall (y,\alpha)\in\LL_{2\,}^r \!\times\! [0,1])
~\big\langle 
(y,u),
\varPi 
(y,u)
\big\rangle \! \geq \! 0,
~
u\!=\!\alpha \varDelta(y).
\end{equation*}
Further, given stable $G:\LL_{2e}^q\rightarrow \LL_{2e}^r$ such that $[\![G,\alpha \varDelta]\!]$ is well-posed for all $\alpha\in[0,1]$, suppose
\begin{equation*}
(\exists \epsilon\in\mathbb{R}_{>0})
(\forall u\in\LL_{2\,}^q)
~\big\langle 
(y,u),
\varPi 
(y,u)
\big\rangle \!\leq\! -\epsilon \| u\|^2\!,
~
y\!=\! G(u).
\end{equation*}
Then, $[\![G,\varDelta]\!]$ is stable. 
\end{theorem}

\subsection{Graphs} \label{sec:graphs}
Let $\G=(\V,\E)$ be a simple (self-loopless and undirected) connected graph, where $\V=[n]$ is the set of $n\in\naturals\setminus\{1\}$ vertices, and $\E\subset\{\{i,j\}\,|\,i\neq j\in\V\}$ is the set of $m:=|\E|\in\naturals$ edges. The elements of $\N_i:=\{j\in\V\,|\,\{i,j\}\in\E\}$ are the neighbours of $i\in\V$, and $\E_i:=\{\{i,j\}\,|\,j\in\mathcal{N}_i\}$ is the corresponding neighbourhood edge set; $j\in\N_i \Leftrightarrow i\in\N_j$ and $m_i:=|\N_i|=|\E_i|\geq 1$ for all $i\in\V$ by connectedness. Given $\F\subset\E$, the corresponding edge-induced sub-graph is $\G[\F]:=(\V_\F,\F)$, where $\V_\F:=\{i\in\V\,|\,\E_i\cap\F\neq\emptyset\}$.

The bijective map $\kappa:\E\rightarrow \M$, with $\M:=[m]$, fixes an enumeration of the edge set $\E$.
Similarly, bijective map
$\kappa_{i}:\N_i\rightarrow \M_i$, with $\M_i:=[m_i]$, fixes an enumeration of the neighbours / edges associated with $i\in\V$. These bijections are 
used to index matrix representations of network structure.

\section{Network modelling for robustness analysis}
\label{sec:ideal+IQC} 

\subsection{\mc{An example network}} \label{subsec:examp}

\mc{For context in subsequent sections, including the numerical example in Section~\ref{sec:numsec}, consider a network of $n\in \naturals\setminus\{1\}$ force actuated agents, each with mass $\mu_i$ and position $x_i$ relative to an associated equilibrium; i.e., $\mu_i\ddot{x}_i=u_i$ for $i\in[n]$. Mass-spring-damper chain dynamics is achieved under the distributed state-feedback law for the force input 
\begin{align*}
& u_i=-\gamma_{i,i} x_{i} \!-\! \sigma_{i,i} \dot{x}_{i} \!+\! \gamma_{i,i-1}(x_{i-1}\!-\!x_{i})\! +\! \sigma_{i,i-1}(\dot{x}_{i-1}\!-\!\dot{x}_{i}) \\
& \qquad\qquad + \gamma_{i,i+1}(x_{i+1}\!-\!x_{i})
\!+\! \sigma_{i,i+1}(\dot{x}_{i+1}\!-\!\dot{x}_{i} )
\end{align*}
given positive parameters $\gamma_{i,i}$, $\sigma_{i,i}$, $\gamma_{i,i-1}=\gamma_{i-1,i}$ and $\sigma_{i,i-1}=\sigma_{i-1,i}$ for $i\in[n]$, and $\gamma_{n,n+1}=\sigma_{n,n+1}=0$, with $x_0=\dot{x}_0=0$ as the spatial boundary conditions. In the frequency domain, the network corresponds to
\begin{align} \label{eq:msdchain}
    \mathcal{L}[x_i](s) = h_{i,i-1}(s) \mathcal{L}[x_{i-1}](s) + h_{i,i+1}(s) \mathcal{L}[x_{i+1}](s)
\end{align}
for $i\in[n]$, where $\mathcal{L}[\cdot]$ denotes the Laplace transform, and 
\begin{align} \label{eq:exampleagentdynamics}
    h_{i,j}(s) \!:=\! \frac{\sigma_{i,j} s + \gamma_{i,j}}{\mu_i s^2 + \sigma_i s + \gamma_i }\quad \text{for } j\in\{i-1,i+1\},
\end{align}
$\sigma_i := \sigma_{i,i} + \sigma_{i,i-1} + \sigma_{i,i+1}$, and $\gamma_i := \gamma_{i,i}+ \gamma_{i,i-1} + \gamma_{i,i+1}$. The corresponding network graph is $\G=(\V,\E)$ with agent vertices $\V=[n]$, and edges $\E=\{\{i,i+1\}~|~i\in[n-1]\}$. The number of edges is $m:=|\E|=n-1$. One can take $\kappa=(e\mapsto\min e)$ to be the edge enumeration. For $i\in\V$,  $m_i:=|\N_i|=|\E_i|$ is $1$ for $i\in\{1,n\}$, and $2$ otherwise.}

\mc{A possible implementation involves each agent $i\in\V$ revealing only $x_i$ as its output. Given~\eqref{eq:msdchain}, this output depends dynamically on inputs from neighbours in $\G=(\V,\E)$. The order of the input coordinates is fixed by the neighbour enumeration $\kappa_i:\N_i\rightarrow\M_i$, with $\M_i=[m_i]$, which can be set according to $(i-1) \mapsto 1$ and $(i+1) \mapsto 2$ for $i\in[2:n-1]$, $\kappa_1(2) = 1$, and $\kappa_n(n-1)=1$, here. }

\mc{Realization of the `links' used to reveal agent positions to neighbours may be non-ideal, yielding different network dynamics to the intended mass-spring-damper chain obtained with ideal links. The directed link from the output $x_i$ of agent $i\in\V$ to the input of neighbour $j\in\N_i$ could exhibit uncertain dynamics, e.g., time-varying non-linear gain or aperiodic sampling induced delay, that differs across neighbour.}

\subsection{\mc{Uncertain-link network modelling and IQC analysis}}

Consider a network of $n\in \naturals\setminus\{1\}$ dynamic agents. Suppose the agents are coupled according to the simple connected graph $\G=(\V,\E)$. The vertex set $\V=[n]$ is a fixed enumeration of the agents. The edge set is such that $\{i,j\}\in\E$ if the output of agent $i\in\V$ depends on input from agent $j\in\V$, or {\em vice versa}. \mc{Define} $m:=|\E|$, $\M:=[m]$, and for each $i\in\V$, $m_i:=|\N_i|=|\E_i|$ and $\M_i:=[m_i]$, \mc{the latter being} non-empty by connectedness.

To tame the notation, each agent $i\in\V$ has a single output with dynamic dependence on inputs from neighbours. The possibly \mc{unbounded but causal} linear dependence \mc{on inputs from neighbours} is modelled by $H_i:\LL_{2e}^{m_i}\rightarrow \LL_{2e}$. The order of input coordinates is fixed 
by the neighbour enumeration $\kappa_i:\N_i\rightarrow\M_i$. \mc{If the output} \mc{of $H_i$ does not depend on input from $j\in\N_i$ because of directed information flow, the corresponding component $H_{i,\kappa_i(j)}$ of $H_i=\mathbf{1}_{1,m_i}(\oplus_{k\in\M_i} H_{i,k})$ can be set to $0$.} The possibly time-varying non-linear uncertain link from the output of agent $i\in\V$ to its neighbour $j=\kappa_i^{-1}(k)\in\N_i$ is \mc{modelled by causal}  $\varLambda_{i,k}:\LL_{2e}\rightarrow \LL_{2e}$, for $k\in\M_i$. 

The proposed network model 
is \mc{given by the feedback interconnection} 
$[\![P, \varLambda \circ (T H)]\!]$,  where 
\begin{subequations}
\label{eq:blkdiag}
\begin{align} 
\varLambda&:={\textstyle \bigoplus_{i\in\V}} \big({\textstyle \bigoplus_{k\in\M_i}} \varLambda_{i,k}\big):\LL_{2e}^{2m}\rightarrow\LL_{2e}^{2m},\\
T&:={\textstyle \bigoplus_{i\in\V}} \mathbf{1}_{m_i,1}:\LL_{2e}^{n}\rightarrow \LL_{2e}^{2m}, \label{eq:Tdef} \\
H&:={\textstyle \bigoplus_{i\in\V}} H_i:\LL_{2e}^{2m}\rightarrow \LL_{2e}^{n}, 
\end{align}
\end{subequations}
and
$P:\LL_{2e}^{2m}\rightarrow \LL_{2e}^{2m}$ is pointwise multiplication by a {\em permutation} matrix structured according to $\G=(\V,\E)$. For $i\!\in\!\V$, $k\!\in\!\M_i$, 
$r\!=\!(\sum_{h\in[i-1]} m_h + k) \!\in\! [2m]$, and $q\!\in\![2m]$, the corresponding entry of this `routing' matrix is given by
\begin{align} \label{eq:permute}
 P_{(r,\,q)}\!=\!
 \begin{cases} 1 & 
 \text{if}~ q=\sum_{h\in[j-1]} m_h + \kappa_j(i)
 \\
  & \qquad\qquad\qquad\qquad \text{with } j=\kappa_i^{-1}(k),\\
  0 & \text{otherwise}.
\end{cases}
\end{align}

\begin{figure}[t]
\begin{minipage}{.48\linewidth}
\centering
\resizebox{2.25cm}{2.25cm}{
\begin{tikzpicture}[main/.style = {draw, circle}] 
\node[main] (1) {$1$}; 
\node[main] (2) [above of=1] {$2$};
\node[main] (3) [below right of=1] {$3$}; 
\node[main] (4) [below left of=1] {$4$};
\draw (1) -- (2);
\draw (1) -- (3);
\draw (1) -- (4);
\draw (4) -- (3);
\end{tikzpicture}} 
\end{minipage}
\hfill
\begin{minipage}{.48\linewidth}
\centering
\resizebox{3cm}{3cm}{
\begin{tikzpicture}[main/.style = {draw, circle}] 

\node[main] (1) {\,}; 
\node[main] (2) [below of=1,yshift=2mm] {\,};
\node[main] (3) [below right of=2,yshift=2mm,xshift=-4mm] {\,}; 
\node[main] (4) [below left of=2,yshift=2mm,xshift=4mm] {\,};
\node[main] (5) [below left of=4, yshift=+2mm] {\,};
\node[main] (6) [below right of=4, yshift=-2mm,xshift=-10mm] {\,};
\node[main] (7) [below left of=3,yshift=-2mm,xshift=10mm] {\,};
\node[main] (8) [below right of=3,yshift=+2mm] {\,};

\node [rotate=43][draw,dashed,inner sep=0pt, circle,yscale=.5, fit={(7) (8)}] {};
\node [draw,dashed,inner sep=0pt, circle,yshift=-1mm,xscale=.9,yscale=.9, fit={(2) (3) (4)}] {};
\node[rotate=-43][draw,dashed,inner sep=0pt, circle,yscale=.5, fit={(5) (6)}] {};

\node [below of=1, xshift=8mm] {$1$};
\node [right of=1, xshift=-6mm] {$2$};
\node [below left  of=8, yshift=3mm, xshift=10mm] {$3$};
\node [below right of=5, yshift=3mm, xshift=-10mm] {$4$};

\draw (1) -- (2);
\draw (4) -- (5);
\draw (3) -- (8);
\draw (6) -- (7);

\end{tikzpicture}}
\end{minipage}
\caption{\mc{A} network graph $\G$ (left) and corresponding $1$-regular sub-system graph $\G^\star = \bigsqcup_{e\in\E} \G[e]$ with adjacency matrix $P$ (right).}  
\label{fig:example_net}                    
\end{figure} 
As defined, $P=P^\prime=P^{-1}$ is the adjacency matrix of the $1$-regular graph 
 $\G^\star:=(\V^\star,\E^\star)$, with $2m$ vertices and $m$ edges, obtained by taking the disjoint union of the two-vertex sub-graphs $\G[e]$ induced by each edge $e\in\E$ of the network graph $\G=(\V,\E)$; see Figure~\ref{fig:example_net}.
In particular,  
\begin{equation}
 \label{eq:Padjacency}
  P=I_{2m}-L
 \end{equation}
\mc{where} the Laplacian $L=L^\prime \in\mathbb{R}^{2m\times 2m}$ decomposes as
\begin{equation}
 \label{eq:Laplacian_Subsystem_Graph}
 L=B B^\prime ={\textstyle \sum_{k\in \M}}B_{(\cdot,k)}B_{(\cdot,k)}^\prime 
\end{equation}
with $B\in\real^{2m\times m}$ defined by $B_{(r,k)}=1=-B_{(\hat{r},k)}$ for $r = \sum_{h\in[i-1]} m_h + \kappa_i(j)$ and $\hat{r}=\sum_{h\in[j-1]} m_h + \kappa_j(i)$ with $\{i,j\} = \kappa^{-1}(k)$, and $B_{(\tilde{r},k)}=0$ for $\tilde{r} \in [2m]\setminus\{r,\hat{r}\}$, over $k\in\M$. This corresponds to an enumeration of $\V^\star$ that aligns with the neighbour enumerations $\kappa_i$ and $\kappa_j$ in~\eqref{eq:permute}.

\mc{Given $d_v\in\LL_{2e}^{2m}$ and $d_w\in\LL_{2e}^{2m}$, the model $[\![P, \varLambda \circ (T H)]\!]$ is defined by the existence of $(v,w)$ such that $v = P w + d_v$ and $w = (\varLambda\circ (TH))(v) + d_w$; cf.~\eqref{eq:feedback1}. Since $P$ is a linear bijection, this
corresponds to the existence of $(\tilde{v},\tilde{w})$ such that
\begin{align} \label{eq:nominal_net}
    \tilde{v} = HP \tilde{w}, \qquad \tilde{w} = (\varLambda \circ T)(\tilde{v}) + d
\end{align}
with $d=d_w+P^{-1}d_v$.} 
\begin{assumption}
\label{ass:nom_stab} 
The \mc{nominal} network \mc{model $[\![P,TH]\!]$} with ideal links (i.e., $\varLambda=I_{2m}$) is stable. 
\end{assumption}

\mc{Assumption~\ref{ass:nom_stab} holds for the  network in Section~\ref{subsec:examp}. It limits admissible agent dynamics as it is necessary $(P - T H) = (I_{2m}-THP)P$ has a stable inverse. On the other hand, it enables the use of a loop transformation to represent links as perturbations of ideal connections for robustness analysis.} \mc{\begin{assumption}
\label{ass:stab_Del_G} 
There exist $S=\bigoplus_{i\in\V}S_i:\LL_{2e}^n\rightarrow\LL_{2e}^n$ and $S^\dagger=\bigoplus_{i\in\V}S_i^\dagger:\DD\subset \LL_{2e}^n\rightarrow \LL_{2e}^n$, both linear, such that 
\begin{align}
    \varDelta:= (\varLambda-I_{2m}) \circ TS 
\quad\text{and}\quad
G :=  S^\dagger H (P - T H )^{-1} 
\label{eq:G}
\end{align}
are stable, and $\varDelta \circ G=(\varLambda-I_{2m}) \circ T H (P - TH)^{-1}$.  
\end{assumption}

Assumption~\ref{ass:stab_Del_G} limits admissible link dynamics. When $\varLambda$ is stable, taking $S=I_n=S^\dagger$ suffices, since in this case, Assumption~\ref{ass:nom_stab} implies $T H P ( I_{2m} - T  H  P)^{-1} = T G$ is stable, whereby $G = (\bigoplus_{i\in\V} \frac{1}{m_i}) T^\prime T G$ is stable. For certain instances of unstable links, appropriate selection of $S$ can yield stable $\varDelta$, e.g., an integrator in the case of sampling induced delay~\cite{mirkin2007some,Cantoni2020}; see Section~\ref{sec:numsec}. The corresponding $S^\dagger$ further limits admissible dynamics, e.g., to systems $H$ having differentiable image in the case a differentiator.}

\begin{theorem} 
\label{thm:stable_LoopTransformed} 
If $[\![G,\varDelta]\!]$ is stable, with $G$ and $\varDelta$ as per \eqref{eq:G}, then \mc{the uncertain network model} $[\![P,\varLambda\circ (T H)]\!]$ is stable. 
\end{theorem}
\begin{proof}
\mc{By hypothesis, $(I_{2m} - \varDelta\circ G)$ is causally invertible. For $d_v,d_w\in\LL_{2e}^{2m}$, let $d=d_w+P^{-1}d_v \in\LL_{2e}^{2m}$ and
\begin{align} \label{eq:ud}
    u = (I_{2m} - \varDelta\circ G)^{-1} d = (I_{2m} - \tilde{\varDelta}\circ\tilde{G})^{-1}d\in\LL_{2e}^{2m},
\end{align}
where in line with Assumption~\ref{ass:stab_Del_G}, $\tilde{\varDelta}:=(\varLambda-I_{2m})\circ T$ and 
\[\tilde{G} := H(P-TH)^{-1} = ({\textstyle \bigoplus_{i\in\V} \frac{1}{m_i}}) T^\prime T HP (I_{2m}-THP)^{-1}.\] 
Note the causal dependence of $u$ on $d$, and the stability of 
$\tilde{G}$ given $T HP (I_{2m}-THP)^{-1}$ is stable by Assumption~\ref{ass:nom_stab}, which also makes $(I_{2m}-THP)^{-1}=P(P-TH)^{-1}$ stable. Let $\tilde{v}=\tilde{G}u$ and $\tilde{w} = (I_{2m}-THP)^{-1}u$. Then, $\tilde{v}=HP\tilde{w}$ and $\tilde{w}=u+T\tilde{v} = d + \tilde{\varDelta}(\tilde{v})+T\tilde{v}=d+(\varLambda\circ T)(\tilde{v})$ from~\eqref{eq:ud}. As such, \eqref{eq:nominal_net} holds with causal dependence of $(\tilde{v},\tilde{w})$ on $u$, and thus, $d$. Hypothesized stability of $[\![G,\varDelta]\!]$ implies $(\exists c_1\in\mathbb{R}_{>0})(\forall d\in\LL_2^{2m}) \|u\| \leq c_1 \|d\|$. 
Furthermore, $(\exists c_2\in\mathbb{R}_{>0})(\forall u\in\LL_2^{2m}) \|(\tilde{v},\tilde{w})\| \leq c_2 \|u\|$ because $\tilde{G}$ and $(I_{2m}-THP)^{-1}$ are stable. Let $v=P\tilde{w}$ and $w=\tilde{w}-P^{-1}d_v$. Then, $v=P w + d_v$ and $w=(\varLambda\circ(TH))(v) + d_w$, with $\|v\| =  \|P\tilde{w}\|=\|\tilde{w}\|\leq c_2\cdot c_1 \cdot \sqrt{2} \|(d_v,d_w)\|$ and $\|w\|\leq \|\tilde{w}\| + \|P^{-1}d_v\| \leq (c_2\cdot c_1 +1)\cdot \sqrt{2} \|(d_v,d_w)\|$ whenever $(d_v,d_2)\in\LL_2^m\times\LL_2^m$, since $P$ is an isometry and $\|d\| \leq \|d_w\|+\|d_v\|\leq \sqrt{2}\|(d_v,d_w)\|$.}
\end{proof}

Towards \mc{characterizing} the \mc{perturbation} 
$\varDelta$ 
in~\eqref{eq:G}, let $\varDelta_i:=(\bigoplus_{k\in\M_i}\varLambda_{i,k}-I_{m_i}) \circ \mathbf{1}_{m_i, 1}\mc{S_i}$ for each $i\in\V$, and suppose the \mc{bounded linear self-adjoint multiplier}
\begin{align*}
\mc{\bar{\varPi}}_{i}^*=\mc{\bar{\varPi}}_{i}:=
\begin{bmatrix} \varPi_{1,i} & \varPi_{2,i} \\ \varPi_{2,i}^*
 & \varPi_{3,i} \end{bmatrix}
:\LL_{2\,}\times\LL_{2\,}^{m_i} \rightarrow 
\LL_{2\,} \times \LL_{2\,}^{m_i}
\end{align*}
(potentially parametrized as in Section~\ref{sec:numsec})
is given such that \mc{the following IQC holds:}
\begin{align} 
&(\forall (y_i,\alpha)\in\LL_2\times [0,1])\nonumber \\
&\qquad\qquad \innerp{(y_i,u_i)}{\mc{\bar{\varPi}}_i(y_i,u_i)} \geq 0,  
\quad u_i=\alpha\varDelta_i(y_i).
\label{eq:DELiIQC}
\end{align}
\mc{Since} $\varDelta = \bigoplus_{i\in\V} \varDelta_i$, it \mc{can be shown} that
\begin{align} 
&(\forall (y,\alpha)\in \LL_{2\,}^{n}\times [0,1])\nonumber \\
&\qquad\qquad \innerp{(y,u)}{\varPi(y,u)} \geq 0, \quad 
u=\alpha \varDelta(y) \label{eq:DeltaIQC}
\end{align}
where
$\varPi=\varPi^*:(\LL_{2\,}^{n} \times \LL_{2\,}^{2m}) \rightarrow (\LL_{2\,}^{n} \times \LL_{2\,}^{2m})$ 
is given by
\begin{align}
\label{eq:multip}
    \varPi = \begin{bmatrix} \varPi_1 & \varPi_2 \\ \varPi_2^* & \varPi_3 \end{bmatrix}
    =\begin{bmatrix} \bigoplus_{i\in\V} \varPi_{1,i} &  \bigoplus_{i\in\V}\varPi_{2,i} \\ \bigoplus_{i\in\V} \varPi_{2,i}^* & \bigoplus_{i\in\V} \varPi_{3,i} \end{bmatrix}.
\end{align}

\begin{theorem} \label{thm:Unstructstabcert}
Given $G$ and $\varPi_{\bullet}$ in \eqref{eq:G} and~\eqref{eq:multip}, suppose 
 \begin{align}
 &(\exists \epsilon\in\mathbb{R}_{>0})~(\forall u \in \LL_{2\,}^{2m})\nonumber \\
 \label{eq:stability_l2+}
 &\quad   	\innerp{\begin{bmatrix}
	  y \\
		u 
        \end{bmatrix}}{\begin{bmatrix} \varPi_1 & \varPi_2 \\ \varPi_2^* & \varPi_3 \end{bmatrix}\begin{bmatrix}
	  y \\
		u 
        \end{bmatrix}} \leq -\epsilon \| u \|^2\!,
        \quad y=Gu.
\end{align}
Then, the uncertain network \mc{model} $[\![P,\varLambda\circ (T H)]\!]$ is stable.
\end{theorem}
\begin{proof}
Stability of $\sys{G}{\varDelta}$ follows from Theorem~\ref{thm:robust_stability} and~\eqref{eq:DeltaIQC}. \mc{Thus}, $[\![P,\varLambda\circ (T H)]\!]$ is stable by Theorem~\ref{thm:stable_LoopTransformed}.
\end{proof}

\begin{remark} \label{rem:unstruct}
In~\eqref{eq:stability_l2+}, the structure of $\varPi_{\bullet} = \bigoplus_{i\in\V} \varPi_{\bullet,i}$ is the same as $\varDelta=\bigoplus_{i\in\V}\varDelta_i$. On the other hand, $G:\LL_{2e}^{2m}\rightarrow \LL_{2e}^n$ typically lacks structure, as $(P-TH)^{-1}$  in~\eqref{eq:G} may \mc{not be sparse}. \mc{This precludes direct application of results from~\cite{khong2014scalable,pates2016scalable}}. In Section~\ref{subsec:structcert}, a coprime factorization \mc{is used to formulate an} equivalent structured robust stability certificate \mc{and overlapping edge-partition based decomposition.}
\end{remark}

\section{Main results} \label{sec:mainresults}

\subsection{An equivalent structured robust stability certificate} \label{subsec:structcert}

\begin{assumption} \label{ass:coprime}
$H_i= N_i D_i^{-1}$ for each agent $i\in\V$, where $N_i:\LL_{2e}^{m_i}\rightarrow \LL_{2e}$ and $D_i:\LL_{2e}^{m_i}\rightarrow\LL_{2e}^{m_i}$ are stable linear systems, and $U_i N_i + V_i D_i = I_{m_i}$ for some stable linear sytems  $U_i:\LL_{2e}\rightarrow\LL_{2e}^{m_i}$ and $V_i:\LL_{2e}^{m_i}\rightarrow\LL_{2e}^{m_i}$; i.e., $H_i=N_i D_i^{-1}$ is a {\em coprime} factorization~\cite{vidyasagar1985control}, \mc{which exists, e.g., whenever the system has a proper rational transfer function.}
\end{assumption}

\mc{It can be shown that}
\begin{subequations}
\label{eq:Gcoprimefactors}
\begin{align} \label{eq:NN} 
        N &:= {\textstyle \bigoplus_{i\in\V}} \mc{S_i^\dagger} N_i \\
        M &:= P\,({\textstyle \bigoplus_{i\in\V}} D_i) - 
        {\textstyle \bigoplus_{i\in\V}} (\mathbf{1}_{m_i,1}  N_i) 
        \label{eq:MM} 
\end{align}         
\end{subequations}
are coprime \mc{and $G=NM^{-1}$. Indeed, with $\tilde{N}:=\bigoplus_{i\in\V}N_i$ and $D:=\bigoplus_{i\in\V}D_i$,  Assumptions~\ref{ass:stab_Del_G}~and~\ref{ass:coprime} make $G=S^\dagger \tilde{N} D^{-1}(P-T\tilde{N}D^{-1})^{-1}$ stable. As such, $N = S^\dagger \tilde{N} = G(PD-T\tilde{N})=GM$ is stable because $M=(PD-T\tilde{N})=(P-TH)D$ is stable. Furthermore, with $U:=\bigoplus_{i\in\V} U_i$ and $V:=\bigoplus_{i\in\V} V_i$, the inverse $M^{-1}=(U \tilde{N} + V D)M^{-1} = U H(P-TH)^{-1} + V(P-TH)^{-1}$ is stable because $\tilde{G}=H(P-TH)^{-1}$ and $(P-TH)^{-1}$ are stable by Assumption~\ref{ass:nom_stab}, as observed in the proof of Theorem~\ref{thm:stable_LoopTransformed}. Finally, $\mathbf{0}_{2m,n} N + M^{-1} M = I_{2m}$, whereby $N$ and $M$ are coprime.}

\mc{\begin{lemma} \label{lem:eqivcpf}
The IQC~\eqref{eq:stability_l2+} is equivalent to 
\begin{align}
&(\exists \epsilon\in\mathbb{R}_{>0})~(\forall z\in\LL_{2}^{2m}) \nonumber \\
\label{eq:stabiliy_G_2}
&\qquad        \innerp{\begin{bmatrix}
	  N \\
		M 
        \end{bmatrix}z}{\begin{bmatrix} \varPi_1 & \varPi_2 \\ \varPi_2^* & \varPi_3 \end{bmatrix}  
        \begin{bmatrix}
	  N \\
		M 
        \end{bmatrix}z}
        \leq -\epsilon \| z \|^2
\end{align}
with $N$ and $M$ as per~\eqref{eq:Gcoprimefactors}.
\end{lemma}}
\mc{\begin{proof}
As noted above, $N$ and $M$ are coprime with $M^{-1}$ stable and $G=NM^{-1}$. Thus, $(y,u)\in\LL_2^{2m}\times\LL_2^n$ satisfies $y=Gu$ if, and only if, there exists $z\in\LL_2^{2m}$ such that $y=Nz$ and $u=Mz$; n.b., if $y=Nz\in\LL_2^{2m}$ and 
$u=Mz\in\LL_2^n$, then $z=\mathbf{0}_{2m,n}y + M^{-1}u\in\LL_2^{2m}$.
As such, \eqref{eq:stability_l2+} is equivalent to \eqref{eq:stabiliy_G_2} because $-\epsilon \|z\|^2 \leq -(\epsilon/\|M\|^2) \|Mz\|^2$ and $-\epsilon \|Mz\|^2 \leq -(\epsilon/\|M^{-1}\|^2) \|z\|^2$ for all $z\in \LL_2^{2m}$.
\end{proof}}

Network structure is apparent \mc{in the equivalent robustness certificate~\eqref{eq:stabiliy_G_2}. Specifically,} $M = J - K$ with 
\begin{align}
\label{eq:JJKK}
J := {\textstyle \bigoplus_{i\in\V}} (D_i-\mathbf{1}_{m_i, 1} N_i)
~\text{ and }~
K := L \, ({\textstyle \bigoplus_{i\in\V}} D_i) 
\end{align}
\mc{where} $L=I-P$
is the sub-system graph Laplacian \mc{in~\eqref{eq:Padjacency}}.
 
The following lemma enables decomposition of~\eqref{eq:stabiliy_G_2} according to edge-based partitions of the network graph,
as pursued in Section~\ref{subsec:partcert}. The result is inspired by the proof of~\cite[Theorem 1]{pates2016scalable} for a somewhat different IQC structure.
\begin{lemma}\label{lem:PVRClusters}  
\mc{Given $c\in\mathbb{N}$}, for each $p\in[c]$ suppose there exist $\epsilon_{p}\in\real_{>0}$, symmetric $W_{p} \succeq \mathbf{0}_{2m,2m}$, bounded linear $X_{p}\!=\!X^*_{p}:\LL_{2}^{2m}\rightarrow\LL_{2}^{2m}$,  $Z_{p}\!=\!Z_{p}^*:\LL_{2}^{2m}\rightarrow\LL_{2}^{2m}$, and $Y_{p}:\LL_{2}^{2m}\rightarrow\LL_{2}^{2m}$,
and stable linear $K_p:\LL_{2e}^{2m}\rightarrow\LL_{2e}^{2m}$, such that $W:={\textstyle{\sum_{p\in[c]}}} W_{p} \succ \mathbf{0}_{2m,2m}$, and 
\begin{subequations}
\label{eq:cluster_generic}
\begin{align}
    \label{eq:cluster_IPk}
       & \innerp{\begin{bmatrix}
		I_{2m} \\
		  K_{p}\\
		\end{bmatrix} z} {\begin{bmatrix}
		X_{p} + \epsilon_{p} W_{p} & Y_{p} \\
    	Y_{p}^{*} & Z_{p}  \end{bmatrix}
     \begin{bmatrix}
		I_{2m} \\
		  K_{p}\\
		\end{bmatrix}z}\leq 0,\\
      &     \innerp{ z}{\varPsi_1 z}  \leq {\textstyle \sum_{p\in[c]}}
      \innerp{z}{ X_{p} z}, 
      \label{eq:cluster_IPX}\\
\label{eq:cluster_IPY}
      &  \innerp{z}{ (K^* \varPsi_2^* \!+\! \varPsi_2 K) z} 
      \leq \textstyle{\sum_{p\in[c]}}\innerp{z}{(K_{p}^*Y_{p}^*\!+\!Y_{p} K_{p}) z}, 
      \\    
\label{eq:cluster_IPZ}
      &     \innerp{z}{ K^* \varPsi_3 K z} \leq {\textstyle \sum_{p\in[c]}}\innerp{z}{K_p^* Z_{p} K_{p} z},
\end{align}   
\end{subequations}
for all $z\in\LL_{2\,}^{2m}$, where
\begin{align} \label{eq:newmult}
    \!\!\begin{bmatrix}
         \varPsi_1 & \varPsi_2 \\
         \varPsi_2^* & \varPsi_3
     \end{bmatrix}
     \!:=\!
     \begin{bmatrix}
         N & \mathbf{0}_{n,2m} \\ J & -I_{2m}
     \end{bmatrix}^*
     \begin{bmatrix}
         \varPi_1 & \varPi_2 \\
         \varPi_2^* & \varPi_3
     \end{bmatrix}
     \begin{bmatrix}
         N & \mathbf{0}_{n,2m} \\ J & -I_{2m}
     \end{bmatrix}\!,
\end{align}
and \mc{$\varPi=\varPi^*$,
$(N,M$), and
$(J,K)$,
are given in~\eqref{eq:multip},~\eqref{eq:Gcoprimefactors}, and~\eqref{eq:JJKK}, respectively}.
Then, 
\mc{IQC~\eqref{eq:stability_l2+} holds}.
\end{lemma}
\begin{proof}
First note~\eqref{eq:stability_l2+} is equivalent to~\eqref{eq:stabiliy_G_2} by Lemma~\ref{lem:eqivcpf}, and for all $z\in\LL_{2}^{2m}$ the constraint in~\eqref{eq:stabiliy_G_2} can be written as
\begin{align}
    \label{eq:stabiliy_G_2b}
        \innerp{\begin{bmatrix}
	  I_{2m} \\
		K 
        \end{bmatrix}z}{\begin{bmatrix} \varPsi_1 + \epsilon I_{2m} & \varPsi_2 \\ \varPsi_2^* & \varPsi_3 \end{bmatrix}  
        \begin{bmatrix}
	  I_{2m} \\
		K
        \end{bmatrix}z}
        \leq 0.
\end{align}
Now, observe that~\eqref{eq:cluster_IPk} implies
$\innerp{z}{X_{p} z} 
    + \innerp{z}{(K_{p}^*Y_{p}^*+Y_{p} K_{p}) z}
    + \innerp{z}{K_{p}^* Z_{p} K_{p} z}
\leq -\epsilon_{p} \innerp{z}{ W_{p} z}$,
$p\in[c]$.
Therefore, 
\begin{align}
& \sum_{p\in[c]}   \Big (\! \innerp{z}{ X_{p} z}  + \innerp{z}{(K_{p}^*Y_{p}^*+Y_{p} K_{p}) z} + \innerp{z}{K_{p}^* Z_{p} K_{p} z} \!\Big ) \nonumber \\ 
&\leq -{\textstyle \sum_{p\in[c]} \epsilon_{p} \innerp{z}{W_p z}} \leq - \epsilon || z ||_2^2, \label{eq:cluster_Proof1}
\end{align}
where $\epsilon = \big(\min_{p\in[c]} \epsilon_{p}\big) \cdot \big(\min_{x^\prime x=1} x^\prime W x\big)>0$ because $W:=\sum_{p\in[c]} W_{p} \succ \mathbf{0}_{2m,2m}$.
Combining \eqref{eq:cluster_IPX}, \eqref{eq:cluster_IPY}, \eqref{eq:cluster_IPZ}, and \eqref{eq:cluster_Proof1}, 
yields~\eqref{eq:stabiliy_G_2b}.
\end{proof} 

\subsection{Decomposition by localized edge-partitions} 
\label{subsec:partcert}

\mc{A cardinality $c\in\naturals$ {\em localized edge-partition} of $\G=(\V,\E)$ 
is any set $\mathscr{P}=\{\F_1,\ldots,\F_c\}\subset 2^{\E}$
such that each sub-graph $\G[\F_{p}]$ is connected, and $\E = \bigcup_{p\in[c]} \F_p$. Connectedness makes partition elements localized.} \mc{For admissible partitions, defined by Assumption~\ref{ass:add_p} below, it is shown how~\eqref{eq:stabiliy_G_2} may be decomposed via Lemma~\ref{lem:PVRClusters} and an apposite selection} of
\begin{itemize}
    \item $W_{p} = W_{p}^\prime\in \real^{2m\times 2m}$, \item 
$X_{p} = X^*_{p}:\LL_{2}^{2m}\rightarrow\LL_{2}^{2m}$, $Z_{p} = Z_{p}^*:\LL_{2}^{2m}\rightarrow\LL_{2}^{2m}$, 
\item $Y_{p}:\LL_{2}^{2m}\rightarrow\LL_{2}^{2m}$,   and $K_p:\LL_{2e}^{2m}\rightarrow\LL_{2e}^{2m}$, 
\end{itemize}
for each $p\in[c]$. The development of this main result is facilitated by \mc{the} additional notation \mc{defined next}.

Given the localized edge-partition $\{\F_1,\ldots,\F_c\}$, 
the associated vertex-partition 
$\{\U_1,\ldots,\U_c\}\subset 2^\V$ \mc{consists of the} naturally ordered edge-induced sub-graph vertex sets
\[\U_p := \{i\in\V\,|\,\E_i\cap\F_p \neq \emptyset\},
  \quad  p\in[c]. \]
\mc{The corresponding set}
$\J_i:= \{p\in[c]\,|\,i\in\U_p\}$ 
collects the indexes of vertex partition elements that contain agent $i\in \V$. 
Since $\bigcup_{p\in[c]} \F_p = \E$ by definition, and since $\G$ is connected, it follows that $\bigcup_{p\in[c]} \U_p = \V$ and $\bigcup_{i\in\V} \J_i=[c]$, whereby
\begin{align} \label{eq:INDEX_SWAP_CV}
\{ (p,i)\,|\,p\in[c],~i\in\U_p \} = \{ (p,i)\,|\,i\in\V,~p\in\J_i \}.
\end{align} 
Further, given the edge-set enumeration $\kappa:\E\rightarrow\M$, the associated partition $\{\K_1,\ldots,\K_c\}\subset 2^\M$ of the edge index set is defined by
\begin{align} \label{eq:KKp}
\K_p:=\{\kappa(e)\,|\,e\in\F_p\} \subset \M,\quad p\in[c].
\end{align}
Also,
$\O_k:=\{p\in[c]\,|\,\kappa^{-1}(k)\in\F_p\} = \{p\in[c]\,|\,k\in\K_p\}$
collects the partition indexes that contain the edge indexed by $k\in\M$. Since $\bigcup_{p\in[c]} \F_p = \E$ by definition, it follows that $\bigcup_{p\in[c]} \K_p = \M$ and $\bigcup_{k\in\M} \O_k = [c]$, whereby
\begin{align} 
\{ (p,k)\,|\,p\in[c],k\in\K_p \} 
\label{eq:INDEX_SWAP_CM}
= \{ (p,k)\,|\,k\in\M,p\in\O_k \}.
\end{align}
Finally: \mc{(i)}
\begin{align} \label{eq:LLk}
\L_k:=\{\kappa(e)\,|\,e\in(\E_i\cup\E_j),~\{i,j\}=\kappa^{-1}(k)\}\big\backslash\{k\} 
\end{align}
collects the edge indexes associated with the two vertices that define the edge indexed by $k\in\M$, excluding the latter; \mc{and (ii)} $\Q_{k,\ell} := \{p\in\O_k\,|\,\ell\in\K_p\}=\{p\in[c]\,|\,\{k,\ell\}\subset\K_p\}$
collects the indexes of the partition elements that contain both $\kappa^{-1}(k)$ and $\kappa^{-1}(\ell)$, for $\ell\in\L_k$. 
\begin{assumption} \label{ass:add_p} The localized edge-partition $\{\F_1,\ldots,\F_c\}$ satisfies
$(\forall k\in\M)~
\bigcup_{p\in\O_k}
\K_p\backslash\{k\} \supset \mathcal{L}_k$.  
\end{assumption}

\mc{
\begin{remark} 
\label{rem:ass2}
Under Assumption~\ref{ass:add_p}, for every $k\in\M$, the union of partitions that contain $\kappa^{-1}(k)\in \E$ covers all edges that share a vertex with it. As such, for all $k\in\M$ and $\ell\in\mathcal{L}_k$, the set $\Q_{k,\ell}$ is non-empty, and 
\begin{align*} 
\{ (p,\ell) \,|\, p\in\O_k,\ell\in\K_p\cap\L_k\}
=
\{ (p,\ell) \,|\, \ell\in\L_k,p\in\Q_{k,\ell}\}.
\end{align*}
Compliant partitions with cardinality $c>1$ must involve overlap; i.e., $(\forall p\in[c])(\exists q\neq p) \F_{p}\cap\F_{q} \neq \emptyset$. Examples are discussed in Remark~\ref{rem:neighbourhoodpart} and Section~\ref{sec:numsec}.
\end{remark}}

\mc{All localized edge-partitions satisfy~\eqref{eq:INDEX_SWAP_CV} and~\eqref{eq:INDEX_SWAP_CM}. With Assumption~\ref{ass:add_p}, the set equality in 
Remark~\ref{rem:ass2} also holds. This defines an admissible set of partitions for the main result, which involves aligned decompositions of the following: }
\begin{itemize}
    \item the sub-system graph Laplacian $L=\sum_{k\in\M}L_k$ in~\eqref{eq:Laplacian_Subsystem_Graph}, with $L_k:=B_{(\cdot,k)}B_{\smash{(\cdot,k)}}^{\smash{\prime}}$; 
    \item the stable system $D:=\bigoplus_{i\in\V} D_i$  in accordance with the \mc{agent} coprime factorizations \mc{in Assumption~\ref{ass:coprime}}; and
    \item the structured multiplier $\varPsi$ in~\eqref{eq:newmult}, which also bears dependence on the \mc{agent coprime} factorizations, as well as the IQC model of the links in~\eqref{eq:DeltaIQC}. 
\end{itemize}
More specifically, for each $p\in[c]$, let 
\begin{subequations} \label{eq:BLpk}
\begin{align}
\hat{B}_{p,k}&:= ({\textstyle\bigoplus_{i\in\U_p}}( \textstyle{\bigoplus_{r\in\I_i}}
B_{(r,k)} ) \mathbf{1}_{m_i, 1}),\quad \mc{k\in\K_p},
\label{eq:Bpk} \\
\hat{L}_{p,k}&:= \hat{B}_{p,k}\hat{B}_{ p,k}^{\prime}, \quad \mc{k\in\K_p},
\label{eq:Lpk} \\
\hat{L}_p &:={\textstyle \sum_{k\in\K_p}}\hat{L}_{p,k}, \label{eq:LLp}\\
   \hat{D}_p &:=\textstyle{\bigoplus_{i\in\U_p}} D_i, \label{eq:DDp}\\
\hat{\varPsi}_{\bullet,p} &:= {\textstyle \bigoplus_{i\in\U_p}} \varPsi_{\bullet,i}\,,
\quad \bullet\in\{1,2,3\},
\label{eq:Psip}
\end{align}
\end{subequations}  
\mc{where 
\begin{align} \label{eq:Ii}
\textstyle \I_i:=[\sum_{h\in[i-1]}m_h+1:\sum_{h\in[i]}m_h], \quad i\in\V.
\end{align}}
\mc{Further}, let
\begin{subequations}
\label{RED_part_Xi}
\begin{align}
\hat{K}_p&:=\tfrac{1}{2}\hat{L}_p \hat{D}_p \, ,
\label{RED_N_AL}\\
\hat{W}_{p} &:=
{\textstyle \bigoplus_{i\in\U_p}} \omega_{i} I_{m_{i}} \, ,
\label{RED_part_CW}\\
\hat{X}_{p} &:= 
{\textstyle \bigoplus_{i\in\U_p}} 
\xi_{i} \varPsi_{1,i} \, ,
\label{RED_part_CX}\\
\hat{Y}_p&:= \hat{\varPsi}_{2,p} \left(
{\textstyle \sum_{k\in\K_p}} \eta_{k}\hat{L}_{p,k}\right), \label{RED_part_Yp}\\
\hat{Z}_p&:= {\textstyle \sum_{k\in\K_p}}\zeta_{k} \hat{L}_{p,k} \hat{\varPsi}_{3,p} \hat{L}_{p,k} \nonumber\\
&\quad~~+ {\textstyle\sum_{k\in \K_p}\sum_{\ell\in \K_p\cap\mathcal{L}_k}}
\theta_{k,\ell} \hat{L}_{p,k} \hat{\varPsi}_{3,p} \hat{L}_{p,\ell} \, ,
\label{RED_part_Zp}
\end{align}
\end{subequations}
with
\begin{subequations}
\label{RED_scalars}
\begin{align}
\omega_{i}&:=\xi_{i}:=1\big/|\J_i|,  &&i\in\V,\label{RED_om_xi}\\
\eta_{k}&:=\zeta_{k}:=1\big/|\O_k|, &&k\in\M,
\label{RED_eta_zeta}\\
\theta_{k,\ell}&:=1\big/|\Q_{k,\ell}|, &&\ell\in\mathcal{L}_k,~k\in\M.
\label{RED_theta}
\end{align}
\end{subequations}

\begin{theorem}\label{prop:partition3}
For each $p\in[c]$, suppose 
\begin{align}
&
(\exists \epsilon_p\in\mathbb{R}_{>0})~(\forall \hat{z}\in\LL_2^{\smash{\hat{m}_p}})\nonumber \\
&~ \innerp{\begin{bmatrix}
		I_{\hat{m}_p} \\
		\hat{K}_p\\
		\end{bmatrix}\hat{z}}{\left[ \begin{array}{cc} \hat{X}_p +\epsilon_p \hat{W}_{p} & \hat{Y}_p \\
    	\hat{Y}_p^* & \hat{Z}_p \end{array} \right]\begin{bmatrix}
            I_{\hat{m}_p} \\
		\hat{K}_p\\
		\end{bmatrix} \hat{z}}\leq 0,
\label{RED_part_IQC}
\end{align}
where \mc{$\hat{m}_p :={\textstyle \sum_{i\in\U_p}} m_i$} and $(\hat{K}_p,\hat{W}_p,\hat{X}_p,\hat{Y}_p,\hat{Z}_p)$ are given in~\eqref{RED_part_Xi}.
Then, the uncertain network $[\![P,\varLambda\circ (T H)]\!]$ is stable.
\end{theorem}
\begin{proof}
By Lemma~\ref{lem:IQC_EXP_3} in the Appendix, given $p\in[c]$ and corresponding $\epsilon_p\in\mathbb{R}_{>0}$,~\eqref{RED_part_IQC} for all $\hat{z}\in\LL_{2\,}^{\smash{\hat{m}_p}}$ if, and only if, 
\begin{align}
\label{iqc:extended_3}
\left\langle 
\begin{bmatrix} I_{2m} \\ K_p \end{bmatrix}z, 
\begin{bmatrix} X_{p}+\epsilon_p W_{p} & Y_p \\ 
Y_p^* & Z_p \end{bmatrix}
\begin{bmatrix} I_{2m} \\ K_p \end{bmatrix}z
\right\rangle \le 0,
\end{align}
for all $z\in\LL_{2\,}^{2m}$, with
\begin{subequations}
\label{eq:pf_prop3}
\begin{align}
\label{pf_prop3:eq_Kp}
&K_p := \tfrac{1}{2}\left({\textstyle \sum_{k\in\K_p}} L_k \right) D \\
&W_{p}:={\textstyle \sum_{i\in\U_p}}\omega_{i}
\big(\textstyle{\bigoplus_{t\in[2m]}} T_{(t,i)}\big)~,
\label{pf_prop3:eq_CW} \\
&X_{p}:=\left({\textstyle \sum_{i\in\U_p}}\xi_{i} 
\big({\textstyle \bigoplus_{t\in[2m]}} T_{(t,i)}\big)\right) \varPsi_{1}~,
\label{pf_prop3:eq_CX} \\
&Y_p :=\varPsi_2 \left({\textstyle \sum_{k\in\K_p}}\eta_{k} L_k\right)~, \label{pf_prop3:eq_Yp}\\
&Z_p := {\textstyle \sum_{k\in\K_p}}\zeta_{k}L_k \varPsi_3 L_k \nonumber \\
& \qquad \qquad  + {\textstyle \sum_{k\in\K_p}} {\textstyle \sum_{\ell\in\K_p\cap\L_k}} \theta_{k,\ell} L_k \varPsi_3 L_{\ell}~,
\label{pf_prop3:eq_Zp}
\end{align}
\end{subequations}
where $\varPsi_{\{1,2,3\}}$ \mc{are} given in~\eqref{eq:newmult}, and $T$ in~\eqref{eq:Tdef}. It follows directly that $W_p\succeq 0$. Further, \eqref{eq:cluster_IPk} holds in view of~\eqref{iqc:extended_3}. 
\mc{In order to apply} Lemma~\ref{lem:PVRClusters},
it remains to verify $W:=\sum_{p\in[c]}W_p \succ \mathbf{0}_{2m,2m}$, and~\eqref{eq:cluster_IPX}--\eqref{eq:cluster_IPZ}. 

\mc{It follows} from~\eqref{pf_prop3:eq_CW} \mc{that}
\begin{align*}
{\textstyle\sum_{p\in[c]}}W_p 
&={\textstyle\sum_{i\in\V}\sum_{p\in\J_i}}\omega_{i}\big(\textstyle{\bigoplus_{t\in[2m]}} T_{(t,i)}\big)\\
&={\textstyle\sum_{i\in\V}}\left(\textstyle{\bigoplus_{t\in[2m]}} T_{(t,i)}\right)=I_{2m} \succ \mathbf{0}_{2m,2m}.  
\end{align*}
The first equality holds by~\eqref{eq:INDEX_SWAP_CV}, and the second because~\eqref{RED_om_xi} implies $\sum_{p\in\J_i}\omega_{i}=1$. Similarly, \mc{it follows} from~\eqref{pf_prop3:eq_CX} \mc{that}
\begin{align*}
{\textstyle \sum_{p\in[c]}} X_p &= 
{\textstyle \sum_{i\in\V} \sum_{p\in\J_i} 
\left( \bigoplus_{t\in[2m]}\xi_{i} T_{(t,i)}\right)} \varPsi_1
=\varPsi_1
\end{align*}
since $\sum_{p\in\J_i}\xi_{i}=1$,
\mc{whereby}~\eqref{eq:cluster_IPX} holds with equality. 

With~\eqref{pf_prop3:eq_Kp} and \eqref{pf_prop3:eq_Yp}, $Y_p K_p=Y_p D$ by Lemma~\ref{lem:tech_1} in the Appendix. 
As such, 
\begin{align*}
{\textstyle\sum_{p\in[c]}}Y_p K_p
&=\varPsi_2  \left(
{\textstyle\sum_{k\in\M}\sum_{p\in\O_k}}
\eta_{k}L_k \right) D\\
&=\varPsi_2  \left(
{\textstyle\sum_{k\in\M}}L_k \right) D
= \varPsi_2 K.
\end{align*}
The first equality holds by~\eqref{eq:INDEX_SWAP_CM}, the \mc{second}
because~\eqref{RED_eta_zeta} implies
$\sum_{p\in\O_k}\eta_{k}=1$, and the last \mc{by}~\eqref{eq:Laplacian_Subsystem_Graph} and~\eqref{eq:JJKK}. Therefore,~\eqref{eq:cluster_IPY} holds with equality. 

Finally, with~\eqref{pf_prop3:eq_Kp} and~\eqref{pf_prop3:eq_Zp}, $K_p^* Z_p K_p = D^*Z_pD$ by Lemma~\ref{lem:tech_1} in the Appendix. \mc{Thus, it follow from}~\eqref{eq:INDEX_SWAP_CM} that
\begin{align}
&{\textstyle \sum_{p\in[c]}} 
K_p^*Z_p K_p \nonumber\\
&=D^*\big({\textstyle \sum_{k\in\M}\sum_{p\in\O_k}}\zeta_{k}L_k \varPsi_3 L_k 
\nonumber \\
&\qquad\quad + {\textstyle \sum_{k\in\M}\sum_{p\in\O_k}\sum_{\ell\in\K_p\cap\mathcal{L}_k}}\theta_{k,\ell} L_k \varPsi_3 L_{\ell}\big)D.\label{eqiv:Zp}
\end{align}
The first term \mc{within the} brackets is equal to $\sum_{k\in\M}L_k \varPsi_3 L_k$ 
since~\eqref{RED_eta_zeta} implies $\sum_{p\in\O_k}\zeta_{k}=1$. \mc{Moreover, by
Remark~\ref{rem:ass2},}
\begin{align*}
&{\textstyle\sum_{k\in\M} \! \sum_{\ell\in\mathcal{L}_k} \!
\sum_{p\in\Q_{k,\ell}}}\theta_{k,\ell} L_k \varPsi_3 L_{\ell} 
\!=\!{\textstyle\sum_{k\in\M} \! \sum_{\ell\in\mathcal{L}_k}} L_k \varPsi_3 L_{\ell},
\end{align*}
where this last equality holds because~\eqref{RED_theta} implies $\sum_{p\in\Q_{k,\ell}}\theta_{k,\ell}=1$. By Lemma~\ref{lem:prop_L} in the Appendix, $\sum_{k\in\M}L_k \varPsi_3 L_k+{\textstyle\sum_{k\in\M} \! \sum_{\ell\in\mathcal{L}_k}} L_k \varPsi_3 L_{\ell}=L\varPsi_3 L$. As such, 
\begin{align*}
{\textstyle \sum_{p\in[c]}}
K_p^*Z_p K_p 
&=D^* L \varPsi_3 LD = K^*\varPsi_3 K, 
\end{align*}
where last equality holds \mc{by}~\eqref{eq:Laplacian_Subsystem_Graph} and~\eqref{eq:JJKK}. Therefore,~\eqref{eq:cluster_IPZ} also holds with equality. 
\end{proof}

\begin{remark} \label{rem:neighbourhoodpart}
    The decomposition considered in~\cite{MCNodes} corresponds to $\F=\{\F_1,\ldots, \F_n\}$ with $n=|\V|$ and $\F_{p}=\E_{p}$, $p\in[n]$. \mc{This is the finest localized edge-partition that satisfies} Assumption~\ref{ass:add_p}. In particular, $|\O_k|=2$ and $\L_k = (\K_i\backslash\{k\})\cup(\K_j\backslash\{k\})$ with $\{i,j\}=\kappa^{-1}(k)$, for all $k\in\M$; \mc{i.e., every edge appears in $2$ edge-partition elements}. The elements of the associated vertex partition satisfy $|\U_p|=1+|\N_p|$, $p\in[n]$. On the other hand, the decomposition originally considered in~\cite{MCLinks} involves the localized edge-partition $\F=\{\F_1,\ldots,\F_m\}$ with  $m=|\E|$ and $\F_{p}=\kappa^{-1}(p)$, $p\in[m]$. The elements of the associated vertex-partition satisfy $|\U_p|=2$, $p\in[m]$. In this case, $Z_p$ in~\eqref{pf_prop3:eq_Zp} does not enable verification of~\eqref{eq:cluster_IPZ}. Instead, \mc{in~\cite{MCLinks},} a very conservative and non-unique diagonal plus semi-definite split of $\varPsi_3$ is used \mc{for this purpose}. 
 \end{remark}

\subsection{\mc{Complexity of Semi-Definite Programming Verification}}

Given \mc{linear time-invariant} state-space models for the coprime factors \mc{in Assumption~\ref{ass:coprime}}, the localized IQC in~\eqref{RED_part_IQC}
can be verified via the Kalman-Yakubovic-Popov (KYP) lemma~\cite{rantzer1996kalman}. \mc{Feasibility of} the resulting linear matrix inequality (LMI) condition can be \mc{in turn tested} via a semi-definite program (SDP). Towards a \mc{KYP-compliant reformulation} of~\eqref{RED_part_IQC}, \mc{with reference to Assumption~\ref{ass:coprime},~\eqref{eq:multip},~\eqref{eq:JJKK},~\eqref{eq:BLpk}, and~\eqref{RED_scalars}, for $p\in[c]$, define $\hat{N}_p:=\bigoplus_{i\in\U_p}\! N_i$,}
\begin{align*}
     \hat{J}_p&:=\hat{D}_p-\hat{T}_p\hat{N}_p=\mc{{\textstyle \bigoplus_{i\in\U_p}} (D_i-\mathbf{1}_{m_i, 1} N_i)}, \\
\hat{\varXi}_p&:= {\textstyle 
(\bigoplus_{i\in\U_p} \mc{\sqrt{\xi_i}}) 
\oplus (\bigoplus_{i\in\U_p} \mc{\sqrt{\xi_i}} I_{m_i})},  \\
\mc{\hat{R}}_p &:= \left(
{\textstyle \sum_{k\in\K_p}} \eta_{k}\hat{L}_{p,k}\right), ~\text{ and }~
\hat{\varPi}_p := 
\begin{bmatrix}
\hat{\varPi}_{1,p} & \hat{\varPi}_{2,p} \\
\hat{\varPi}_{2,p}^* & \hat{\varPi}_{3,p}
\end{bmatrix},
\end{align*}
where $\hat{T}_p:=\bigoplus_{i\in\U_p} \mathbf{1}_{m_i,1}$, and $\hat{\varPi}_{\bullet,p}:=
\bigoplus_{i\in\U_p} \varPi_{\bullet,i}$. Then, \mc{recalling~\eqref{eq:newmult}, ~\eqref{RED_part_Xi}, and 
that $N=\bigoplus_{i\in\V} S_i^{\smash{\dagger}} N_i$ is stable as noted below~\eqref{eq:NN}, it follows that}
\begin{align*}
    \hat{Y}_p &= -\begin{bmatrix} (\mc{\hat{S}_p^\dagger} \hat{N}_p)^* 
& \hat{J}_p^* \end{bmatrix}
\begin{bmatrix} \hat{\varPi}_{2,p}^* 
& \hat{\varPi}_{3,p}^* \end{bmatrix}^* 
\mc{\hat{R}}_p, \\
\hat{X}_p &= 
\begin{bmatrix} (\mc{\hat{S}_p^\dagger}\hat{N}_p)^* 
& \hat{J}_p^* \end{bmatrix}
\hat{\varXi}_p \hat{\varPi}_p \hat{\varXi}_p
\begin{bmatrix} (\mc{\hat{S}_p^\dagger} \hat{N}_p)^* 
& \hat{J}_p^* \end{bmatrix}^*,
\end{align*}
\mc{where $\hat{S}_p^\dagger := \bigoplus_{i\in\U_p}S_i^\dagger$.} 
\mc{As such, since} $\hat{Y}_p\hat{K}_p=\hat{Y}_p\hat{D}_p$ and $\hat{K}_p^*\hat{Z}_p\hat{K}_p = \hat{D}_p^* \hat{Z}_p \hat{D}_p$ by Lemma~\ref{lem:tech_1} \mc{in the Appendix}, \mc{the IQC \eqref{RED_part_IQC} is equivalent to the
  existence of $\epsilon_p\in\mathbb{R}_{>0}$ such that}
\mc{\begin{align} \label{eq:KYPform}
\innerp{\begin{bmatrix}
    \hat{G}_p \\ I_{\hat{m}_p}
\end{bmatrix}\!\hat{z}}{\begin{bmatrix}
    \hat{\varUpsilon}_p & \mathbf{0}_{2n_p\!+\hat{m}_p,\hat{m}_p}\\
    \mathbf{0}_{\hat{m}_p,2n_p\!+\hat{m}_p} & 
    \epsilon_p \hat{W}_p
\end{bmatrix} 
\begin{bmatrix}
    \hat{G}_p \\ I_{\hat{m}_p}
\end{bmatrix}\! \hat{z}} \leq 0
\end{align}}
for all $\hat{z}\in\LL_2^{\smash{\hat{m}_p}}$, 
\mc{where $n_p:=|\U_p|$, $\hat{m}_p:=\sum_{i\in\U_p}m_i$,
\begin{align}
\label{eq:Ghat}
    \hat{G}_p &:= 
    \begin{bmatrix}
        (\hat{S}_p^\dagger\hat{N}_p)^* & \hat{N}_p^* & \hat{D}_p^*
    \end{bmatrix}^*,\\
    \hat{\varUpsilon}_p
    &:=
\mc{\hat{Q}}_p^*
\begin{bmatrix}
\hat{\varXi}_p \hat{\varPi}_p \hat{\varXi}_p & \begin{bmatrix} \hat{\varPi}_{2,p}
\\ \hat{\varPi}_{3,p} \end{bmatrix}\!\mc{\hat{R}}_p \\
\mc{\hat{R}}_p^*\begin{bmatrix} \hat{\varPi}_{2,p}^* 
& \hat{\varPi}_{3,p}^* \end{bmatrix} & \hat{Z}_p
\end{bmatrix}\mc{\hat{Q}}_p, \nonumber\\
\nonumber
\hat{Q}_p &:= 
    \begin{bmatrix}
        \begin{bmatrix}
            I_{n_p} \\
            \mathbf{0}_{\hat{m}_p,n_p} 
        \end{bmatrix}
        & \begin{bmatrix}
          \mathbf{0}_{n_p,n_p} \\
          -\hat{T}_p
        \end{bmatrix}
        & \begin{bmatrix}
              \mathbf{0}_{n_p,\hat{m}_p} \\
          I_{\hat{m}_p}
        \end{bmatrix} \\
        \mathbf{0}_{\hat{m}_p,n_p} & \mathbf{0}_{\hat{m}_p,n_p} & -I_{\hat{m}_p}
    \end{bmatrix}.
\end{align}
For static multipliers $\hat{\varPi}_{\bullet,p}$, if $S^\dagger$ is such that $\hat{G}_p$ admits a state-space realization, then the standard KYP lemma applies. Multiplier parametrization is readily accommodated, e.g., to combine different instances of~\eqref{eq:DELiIQC}, or to reflect the common source of the links from each agent. Keeping multiplier parameters free in verification of the resulting LMI reduces conservativeness. Free multiplier variables shared across partition elements lead to coupling of the LMIs. The choice of partition impacts the coupling sparsity.}

\mc{The KYP--LMI for~\eqref{eq:KYPform} has
$\tilde{n}_{p} := \frac{1}{2}\big(\hat{n}^{2}_{p} + \hat{n}_{p}\big) + \hat{o}_p + 1$ 
variables, where $\hat{n}_p:={\textstyle\sum_{i\in\U_p} \nu_i}$ and $\hat{o}_p={\textstyle\sum_{i\in\U_p} o_i}$, with $\nu_i$ as the state dimension of a minimal realization of $((S_i^\dagger N_i),N_i,D_i):\LL_{2e}^{m_i}\rightarrow\LL_{2e}\times\LL_{2e}\times\LL_{2e}^{m_i}$, and $o_i$ the number of multiplier variables in the parametrization of~\eqref{eq:DELiIQC}. The $+1$ term corresponds to the variable $\epsilon_p$. The LMI dimension is $\tilde{m}_p\times\tilde{m}_p$, where $\tilde{m}_p:=(\hat{m}_p+\hat{n}_p)$. }

\mc{For each $p\in[c]$, let the associated LMI variables be grouped into two vectors $\mathsf{v}_p$ and $\mathsf{w}_p$. In particular, $\mathsf{v}_p$ comprises the scalars associated with $\epsilon_p$ in~\eqref{eq:KYPform}, the $\frac{1}{2}\big(\hat{n}^{2}_{p}+\hat{n}_p)$ entries of the symmetric KYP matrix, and the multiplier variables for edges that only appear in $\F_p$. All remaining shared multiplier variables go into $\mathsf{w}_p$. Let $\mathsf{w}$ be the vector of {\em distinct} shared multiplier variables, and let $\mathsf{v}:=(\mathsf{v}_1,\ldots,\mathsf{v}_{c})$. Then, the structured LMI for KYP-based verification of the robustness certificate in Theorem~\ref{prop:partition3} takes the form
\begin{align} \label{eq:structureLMI}
F(\mathsf{v},\mathsf{w}) 
:= {\textstyle \bigoplus_{p\in[c]} } F_p(\mathsf{v}_p,\mathsf{w}) \succeq 0,
\end{align}
where 
\begin{align*}
F_p(\mathsf{v}_p,\mathsf{w})
:= \big(F_{p,0} + \!\!\! {\displaystyle \sum_{i\in [\,|\mathsf{v}_p|\,]} } (\mathsf{v}_{p})_{(i,1)} F_{p,i} + \!\!\! {\displaystyle \sum_{r\in[\,|\mathsf{w}|\,]}} \mathsf{w}_{(r,1)} E_{p,r} \big),
\end{align*}
$F_{p,i}=F_{p,i}^\prime, E_{p,r}=E_{p,r}^\prime \!\in\! \mathbb{R}^{\tilde{m}_p\times\tilde{m}_p}$, $E_{p,r}\!=\!\mathbf{0}_{\tilde{m}_p,\tilde{m}_p}$ if $\mathsf{w}_{(r,1)}$ is not an entry in $\mathsf{w}_p$, and $|\cdot|$ denotes dimension. }

\mc{Testing the feasibility of \eqref{eq:structureLMI} involves solving a SDP~\cite{KaoMeg2007}. The Newton step of an interior-point method for the standard logarithmic barrier, $f(\mathsf{v},\mathsf{w}) = - \log \det F(\mathsf{v},\mathsf{w})$, would involve computation of the gradient, the Hessian and its inverse. Following the analysis in~\cite{KaoMeg2007}, as elaborated in~\cite{CanKao2026}, gradient complexity is  $O(\sum_{p\in[c]} (|\mathsf{v}_p|+|\mathsf{w}|)\tilde{m}_p^3)$, and Hessian complexity is  $O(\sum_{p\in[c]}\tilde{m}_p^2(|\mathsf{v}_p|^2 + |\mathsf{v}_p||\mathsf{w}|+|\mathsf{w}|^2))$. Further,  the Hessian permutes to a block arrowhead structure, with the shared multiplier variables at the head, whereby subsequent computation of the Newton direction $-(\nabla^2 f(\mathsf{v},\mathsf{w}))^{-1} \nabla f(\mathsf{v},\mathsf{w})$ is possible with complexity  $O(|\mathsf{w}|^3+\sum_{p\in[c]}|\mathsf{v}_p|^3)$.
Defining
\begin{align}
&\chi:=|\mathsf{w}|^3 \!+\! 
{\textstyle \sum_{p\in[c]}}  \big(|\mathsf{v}_p|^3 +
\tilde{m}_p^2(|\mathsf{v}_p|^2 
+ |\mathsf{v}_p||\mathsf{w}|+|\mathsf{w}|^2) \nonumber \\
& \qquad \qquad \qquad \qquad \qquad \quad + 
\tilde{m}_p^3(|\mathsf{w}|+|\mathsf{v}_p|) \big)
\label{eq:chi}
\end{align}
the overall complexity is $O(\chi)$ per Newton step.}

\section{Numerical example and conclusion}
\label{sec:numsec}

\mc{Consider~\eqref{eq:msdchain} given $\mu_i=1$, $\gamma_{i,i}=1.399$, $\sigma_{i,i}=0.6528$, $\gamma_{i,i-1}=0.8005$, $\sigma_{i,i-1}=0.3736$, $i\in\V=[40]$. With ideal links, this mass-spring-damper chain is stable, as required in  Assumption~\ref{ass:nom_stab}. Let $\kappa:\E\rightarrow[39]$ and $\kappa_i:\N_i\rightarrow\M_i\subseteq\{1,2\}$ be the edge and neighbour enumerations in  Section~\ref{subsec:examp}. The component $H_{i,k}$ of agent $i\in\V$ corresponds to the stable transfer function in~\eqref{eq:exampleagentdynamics} with $j=\kappa_{i}^{-1}(k)$, for $k\in\M_i$. State-space realizations of $H_i$ and the coprime factors in Assumption~\ref{ass:coprime} can be obtained by standard methods~\cite{vidyasagar1985control}. }

\mc{Sample-and-hold realization of the links that reveal the position $x_i$ of agent $i\in\V$ to its neighbours $j\in\N_i$, makes each $\varLambda_{i,\kappa_i(j)}$ an aperiodic saw-tooth delay that resets to zero at update events~\cite{fridman2004robust}. While this is not stable, the corresponding integrator weighted perturbation from the ideal link is stable. For any bound $h$ on the sampling interval, with $S_i=(v \mapsto (t \mapsto \int_{0}^t v(\tau) \mathrm{d}\tau))$, the $\LL_2$-gain of $(\varLambda_{i,k} - 1)S_i$ is bounded by $2h/\pi$~\cite{mirkin2007some}. A larger bound holds in the case of non-zero residual delay after hold updates~\cite{Cantoni2020}. By taking $S_i^\dagger$ to be differentiation, Assumption~\ref{ass:stab_Del_G} holds. Further, in this setting, each $\hat{G}_p$ in~\eqref{eq:Ghat} admits a linear time-invariant state-space realization for all possible localized edge-partitions. }

\mc{Consider a uniform bound $h$ on the aperiodic sampling interval of all links. This does not presume identical links. Nor does it require  synchronization. For $\gamma:=2h/\pi$, the perturbation of the links serving agent $i\in\V$, given by $\varDelta_i=(\bigoplus_{k\in\M_i} \varLambda_{i,k} - 1)\mathbf{1}_{m_i,1}S_i$, satisfies the IQC~\eqref{eq:DELiIQC} with
\begin{align} \label{eq:examp_multips}
\bar{\varPi}_{i} \!=\!
\begin{bmatrix}
\gamma^2 (\varphi_{1i} + \varphi_{2i}) & 0 &  0 \\
0 & -\varphi_{1i} & 0 \\
0 & 0 & -\varphi_{2i}   
\end{bmatrix}
\end{align}
for $i\in[2,n-1]$, and $\bar{\varPi}_{i}=[\begin{smallmatrix} \gamma^2 \varphi_{1i} & 0\\
 0 & -\varphi_{1i}\end{smallmatrix}]$ for $i\in\{1,n\}$, where $\varphi_{1i} \in\mathbb{R}_{\geq 0}$, $\varphi_{2i} \in\mathbb{R}_{\geq 0}$, are free variables. It turns out that considering the passivity of $\varDelta_i$~\cite{Cantoni2020}, via additional free variables, does not reduce conservativeness in this case.} 

\begin{figure}[b]
    \centering
    \includegraphics[width=0.95\linewidth]{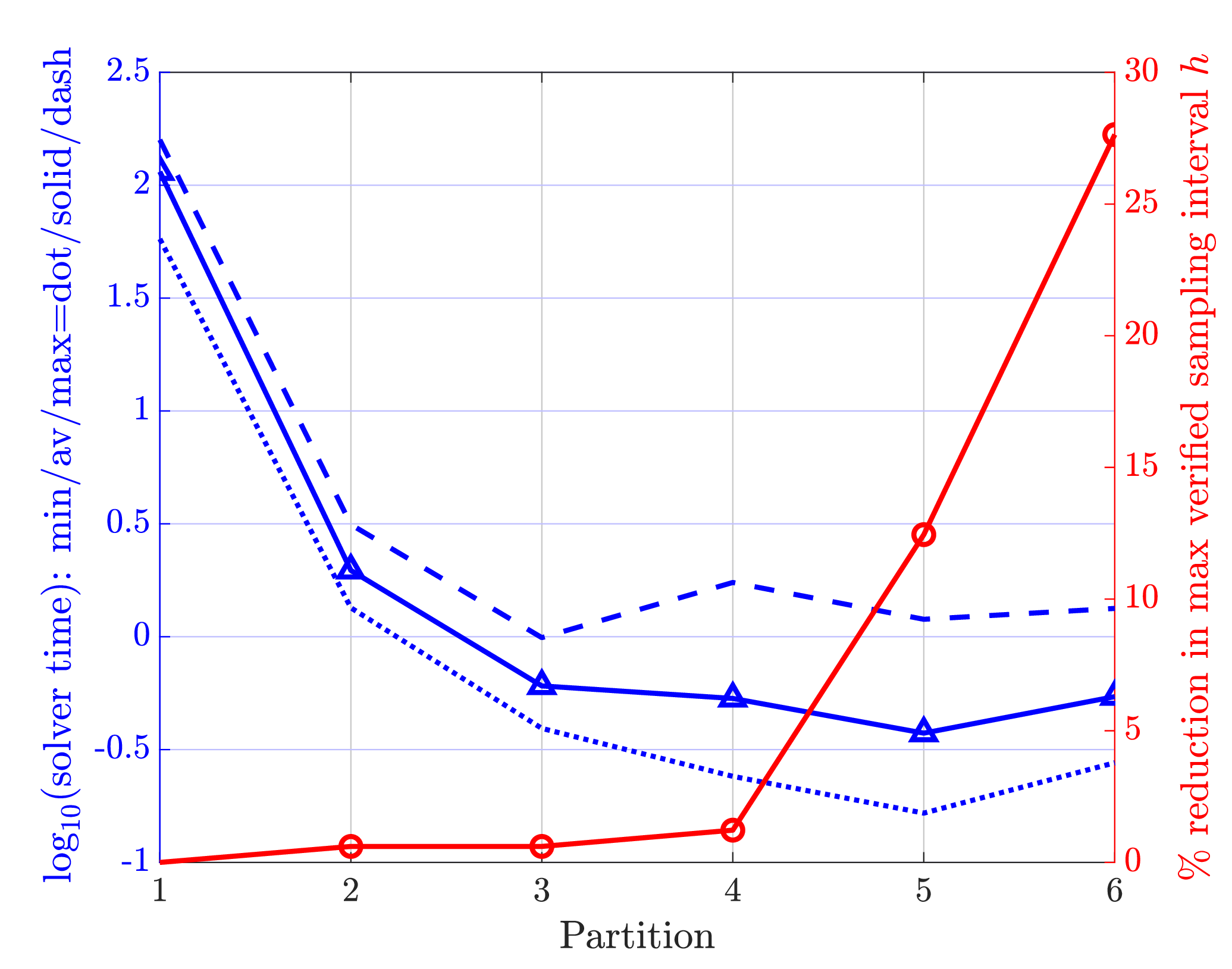}
    \caption{Log solver time and percentage reduction in maximum verified sampling interval relative to the value $h=0.563$ verified for $\mathscr{P}_1$.}
    \label{fig:results_SDex}
\end{figure}

\mc{For the network at hand, the six localized edge-partitions in Table~\ref{tab:partitions} all satisfy Assumption~\ref{ass:add_p} with one edge overlap. Theorem~\ref{prop:partition3} is applied for each partition. In particular, for the parametrized IQC multipliers~\eqref{eq:examp_multips}, SDPs are solved to test the feasibility of corresponding LMIs~\eqref{eq:structureLMI} in a bisection search for the largest verifiable sample interval $h$ in each case. The relevant LMI quantities $|\mathsf{w}|$, $|\mathsf{v}_p|$, $\tilde{m}_p$, and verification complexity estimate $\chi$, are also given in Table~\ref{tab:partitions}. }

\mc{The bisection search involves 15 SDPs for each partition. The YALMIP~\cite{Lofberg2004} parser and SeDuMi~\cite{sturm1999using} solver are used. Figure~\ref{fig:results_SDex} shows minimum, average, and maximum solver time for bisection steps as indicators of verification complexity. The percentage reduction in maximum verified sampling interval from the value obtained using partition $\mathscr{P}_1$ is also shown as a measure of conservativeness. It is evident that complexity decreases with partition refinement. This comes at the cost of a more conservative sampling interval certificate. For the coarse partitions $\mathscr{P}_{\{2,3\}}$ a small conservativeness penalty is incurred for orders of magnitude reduction in computation. Further refinement yields diminished benefit, and overly conservative outcomes with $\mathscr{P}_{\{5,6\}}$. Similar results are reported in~\cite{CanKao2026} for heterogenous agents and asymmetric link uncertainty.}

\mc{In conclusion, the results show that Theorem~\ref{prop:partition3} allows for balancing conservativeness against verification complexity in the analysis of networks with uncertain links. Characterizing the observed transition from significant benefit to diminished returns when refining the edge-partition of the network graph is a challenging direction for future work.} 

\begin{table*}[t]
\centering
\mc{
\begin{tabular}{c|c|c|c|c|c|c|c|c|c|c|}
  & $|\mathsf{w}|$ & \multicolumn{8}{c|}{ $\F_p$, $|\mathsf{v}_p|$, $\tilde{m}_p$ }  & $ \chi\cdot10^{-9}$\\
\hline 
$\!\!\!\!\mathscr{P}_1\!\!\!\!$  & 0 & $\!\!\!e_{[1:39]}\!,\! 3319,\!158\!\!\!$ & & & &  & & & & 324.650\\ 
\hline 
\hline 
$\!\!\!\!\mathscr{P}_2\!\!\!\!$
& 8 & $\!\!\!e_{[1:14]}\!,\! 491,\!59\!\!\!$ & $\!\!\!e_{[14:26]}\!,\! 427,\!56\!\!\!$ & $\!\!\!e_{[26:39]}\!,\! 491,\!59\!\!\!$ & & & & & &2.885\\ 
\hline 
\hline 
$\!\!\!\!\mathscr{P}_3\!\!\!\!$
& 20 & $\!\!\!e_{[1:7]}\!,\! 148,\!31\!\!\!$ & $\!\!\!e_{[7:13]}\!,\! 145,\!32\!\!\!$ & $\!\!\!e_{[13:20]}\!,\! 182,\!36\!\!\!$ & $\!\!\!e_{[20:27]}\!,\! 182,\!36\!\!\!$ & $\!\!\!e_{[27:33]}\!,\! 145,\!32\!\!\!$ & $\!\!\!e_{[33:39]}\!,\! 148,\!31\!\!\!$ & & & 0.259\\ 
\hline
\hline
$\!\!\!\!\mathscr{P}_4\!\!\!\!$  
& 44 & $\!\!\!e_{[1:4]}\!,\! 61,\!19\!\!\!$ & $\!\!\!e_{[4:7]}\!,\! 58,\!20\!\!\!$ & $\!\!\!e_{[7:10]}\!,\! 58,\!20\!\!\!$ & $\!\!\!e_{[10:13]}\!,\! 58,\!20\!\!\!$ & $\!\!\!e_{[13:16]}\!,\! 58,\!20\!\!\!$ & $\!\!\!e_{[16:20]}\!,\! 83,\!24\!\!\!$ & $\!\!\!e_{[20:24]}\!,\! 83,\!24\!\!\!$ & $\!\!\!e_{[24:27]}\!,\! 58,\!20\!\!\!$ 
& 0.060 \\ 
\hline 
\multicolumn{2}{c|}{}  & $\!\!\!e_{[27:30]}\!,\! 58,\!20\!\!\!$ & $\!\!\!e_{[30:33]}\!,\! 58,\!20\!\!\!$ & $\!\!\!e_{[33:36]}\!,\! 58,\!20\!\!\!$ & $\!\!\!e_{[36:39]}\!,\! 61,\!19\!\!\!$ & & & & & \\
\hline
\hline 
$\!\!\!\!\mathscr{P}_5\!\!\!\!$ 
& 72 & $\!\!\!e_{[1:3]}\!,\! 40,\!15\!\!\!$
& \multicolumn{6}{c|}{$\F_p = e_{[2p-1:2p+1]}$,
\ \ $|\mathsf{v}_p|=37$, \ \ $\tilde{m}_p=16$, \ $p=2,\cdots,18$} 
& $\!\!\!e_{[37:39]}\!,\! 40,\!15\!\!\!$ 
& 0.054\\
\hline
\hline 
$\!\!\!\!\mathscr{P}_6\!\!\!\!$ 
& 76  & $\!\!\!e_{[1:2]}\!,\! 23,\!11\!\!\!$
& \multicolumn{6}{c|}{$\F_p = \{e_p,e_{p+1}\}$,
\ \  $|\mathsf{v}_p|=22$, \  \ $\tilde{m}_p=12$, \ $p=2,\cdots,37$} 
& $\!\!\!e_{[38:39]}\!,\! 23,\!11\!\!\!$ & 0.050\\
\hline
\end{tabular}}
\caption{\mc{Partitions $\mathscr{P}_{\bullet}$ and corresponding $|\mathsf{w}|$, $(\F_p, |\mathsf{v}_p|,\tilde{m}_p)$, $p\in[\,|\mathscr{P}_\bullet|\,]$, where $e_{[i:j]}:= \{e_l~|~l\in[i:j]\}$ with $e_l:=\{l,l+1\}$. The scaled Newton step complexity
$\chi \cdot 10^{-9}$ is calculated according to~\eqref{eq:chi} and rounded up to the third decimal place.} }
\label{tab:partitions}
\end{table*}

\appendix

\begin{lemma}
\label{lem:tech_1}
For $k\in\M$, let $L_k = B_{(\cdot,k)}B_{(\cdot,k)}^\prime$ in~\eqref{eq:Laplacian_Subsystem_Graph}. Then,
$L_k  L_{\ell} = 2 L_k$ if $\ell=k$, and 
$\mathbf{0}_{2m,2m}$ otherwise. Further, if $\ell\in\M\backslash(\L_k\cup\{k\})$, with $\L_k$ as defined in~\eqref{eq:LLk}, then $L_k \varPhi L_{\ell}=\mathbf{0}_{2m,2m}$ for every linear $\varPhi=\bigoplus_{i\in\V} \varPhi_i$ with $\varPhi_i:
\LL_2^{m_i}\rightarrow\LL_2^{m_i}$.
\end{lemma}
\begin{proof}
Direction calculation \mc{yields the result} as observed in~\cite{MCLinks}. In particular, $B_{\smash{(\cdot,k)}}^{\smash{\prime}} B_{(\cdot,\ell)}=2$ if $k=\ell$, and $0$ otherwise. Further, $L_k \varPhi L_\ell = B_{(\cdot,k)}(\bigoplus_{i\in\V} \mc{\tilde{B}_{i,k}^\prime} \Phi_i \mc{\tilde{B}_{i,\ell}}) B_{(\cdot,\ell)}^\prime$, where
\begin{align*}
\mc{\tilde{B}_{i,q} = ({\textstyle \bigoplus_{r\in\I_i}} B_{(r,q)})
\mathbf{1}_{m_i,1}, \quad q\in\{k,\ell\},}
\end{align*}
\mc{and $\I_i$ is given in~\eqref{eq:Ii}.}
Since $\G^\star=(\V^\star,\E^\star)$ is $1$-regular, $\ell\notin(\L_k\cup\{k\})$ implies at least one of \mc{$\tilde{B}_{i,k}$} or \mc{$\tilde{B}_{i,\ell}$} is zero, and thus, $L_k \Phi L_\ell = \mathbf{0}_{2m,2m}$.
\end{proof}

\begin{lemma}
\label{lem:prop_L}
Let $\varPhi=\bigoplus_{i\in\V} \varPhi_i$, 
with linear $\varPhi_i:
\LL_2^{m_i}\rightarrow\LL_2^{m_i}$,  
and $L=\sum_{k\in\M}L_k$ as per ~\eqref{eq:Laplacian_Subsystem_Graph}. Then,
$L \varPhi L = {\textstyle \sum_{k\in \M} L_k  \varPhi L_k + \sum_{k\in \M} \sum_{\ell \in \L_k} L_k \varPhi L_\ell ,}$
with $\L_k$ as per~\eqref{eq:LLk}.
\end{lemma}
\begin{proof}
By Lemma~\ref{lem:tech_1}, $L_k \Phi L_\ell = \mathbf{0}_{2m,2m}$ for all $k\in\M$ and $\ell\in\M\backslash(\L_k\cup\{k\})$. Therefore,
\begin{align*}
L\varPhi L 
&={\textstyle \sum_{k\in\M}} L_k \varPhi \big( L_k + {\textstyle \sum_{\ell\in\L_k} L_\ell + \sum_{\ell\in\M\backslash(\L_k\cup\{k\})} L_\ell}\big)\\
&= {\textstyle \sum_{k\in \M} L_k  \varPhi L_k + \sum_{k\in \M} \sum_{\ell \in \L_k} L_k \varPhi L_\ell},
\end{align*}
as claimed.
\end{proof}

\begin{lemma}
\label{lem:tech_2}
For $p\in[c]$ and $k\in\K_p$, with $\K_p$ as per~\eqref{eq:KKp}, let $\hat{L}_{p,k}= \hat{B}_{p,k}(\hat{B}_{p,k})^{\prime}$ as in~\eqref{eq:BLpk}. Then,
$\hat{L}_{p,k} \hat{L}_{p,\ell} = 2 \hat{L}_{p,k}$ for $\ell=k$, and 
$\mathbf{0}_{\hat{m}_p,\hat{m}_p}$ otherwise.
Further, for $\K_{p,1},\ \K_{p,2} \subseteq \K_p$,
${\textstyle \big(\sum_{k\in\K_{p,1}}\hat{L}_{p,k}\big) \big(\sum_{\ell\in\K_{p,2}}\hat{L}_{p,\ell}\big)}
= {\textstyle 2\sum_{k\in\K_{p,1}\cap\K_{p,2}} \hat{L}_{p,k}}$.
\end{lemma}
\begin{proof}
By definition, $\hat{B}_{p,k}^\prime\hat{B}_{p,\ell}=B_{(\cdot,k)}^\prime B_{(\cdot,\ell)}$. \mc{As such}, the result holds by Lemma~\ref{lem:tech_1}.
\end{proof}

\begin{lemma}
\label{lem:IQC_EXP_3}
Given $p\in[c]$ and $\epsilon_p\in\mathbb{R}_{>0}$, the following holds: $\eqref{RED_part_IQC} \text{ for all } \hat{z}\in\LL_{2}^{\smash{\hat{m}_p}} 
~\Leftrightarrow~
\eqref{iqc:extended_3} \text{ for all } z\in\LL_{2\,}^{2m}$.
\end{lemma}
\begin{proof}
Recalling~\eqref{RED_part_Xi}, $\hat{Y}_p\hat{K}_p=\hat{Y}_p\hat{D}_p$ and $\hat{K}_p\hat{Z}_p\hat{K}_p=\hat{D}_p^*\hat{Z}_p\hat{D}_p$ by Lemma~\ref{lem:tech_2}. As such,
\eqref{RED_part_IQC} is $\innerp{\hat{z}}{\smash{\hat{\varXi}_p \hat{z}}} \leq 0$, where 
\begin{align*}
\hat{\varXi}_p &:= {\textstyle \bigoplus_{i\in\U_p}} \big( 
\xi_{i} \hat{\varPsi}_{1,i}
\!+\! \epsilon_p \omega_{i} I_{m_{i}}\big)
 \!+\! \hat{\varPsi}_{2,p}\big( {\textstyle \sum_{k\in\K_p}} \eta_{k}\hat{L}_{p,k}\big)\hat{D}_p \\
 & \qquad + \big(\hat{\varPsi}_{2,p}\big( {\textstyle \sum_{k\in\K_p}} \eta_{k}\hat{L}_{p,k}\big)\hat{D}_p\big)^* + \hat{D}_p^*\hat{Z}_p\hat{D}_p.
\end{align*}
Similarly, recalling~\eqref{pf_prop3:eq_Kp}, $Y_pK_p=Y_pD_p$ and $K_pZ_pK_p=D_p^*Z_pD_p$ by Lemma~\ref{lem:tech_1},~and thus,
\eqref{iqc:extended_3} is $\innerp{z}{\smash{\varXi_p z}} \leq 0$, where
\begin{align*}
&\varXi_p =  {\textstyle \sum_{i\in\U_p}}
\Big(\xi_{i} 
\big({\textstyle \bigoplus_{t\in[2m]}} T_{(t,i)}\big)\varPsi_{1}+\epsilon_p\omega_{i}
\big(\textstyle{\bigoplus_{t\in[2m]}} T_{(t,i)}\big)\Big) \\ 
&~ + \varPsi_2  \big({\textstyle \sum_{k\in\K_p}}\eta_{k} L_k\big) D + \big(\varPsi_2  \big({\textstyle \sum_{k\in\K_p}}\eta_{k} L_k\big) D\big)^* +D^* Z_p D.
\end{align*}
Given $p\in[c]$, and $k\in\K_p$,   
$B_{((s(i-1)+1):s(i),k)}$ in~\eqref{eq:BLpk} is non-zero only if $i\in\U_p$. Therefore, there exists a permutation $\varOmega_p=\varOmega_p^\prime=\varOmega_p^{-1}\in\mathbb{R}^{2m\times 2m}$ for which
$\varOmega_p B_{(\cdot,k)} = \big(\hat{B}_{p,k} \oplus \mathbf{0}_{2m-\hat{m}_p}\big)\mathbf{1}_{2m,1}$,
whereby
$\varOmega_p L_k \varOmega_p' = \hat{L}_{p,k}\oplus \mathbf{0}_{2m-\hat{m}_p,2m-\hat{m}_p}$.
One can then show by direct calculation that $\varOmega_p\varXi_p\varOmega_p^\prime = \hat{\varXi}_p\oplus 
\mathbf{0}_{2m-\hat{m}_p,2m-\hat{m}_p}$, \mc{whereby} the claimed equivalence \mc{holds}. 
\end{proof}

\section*{References}
\vspace*{-20pt}
\bibliography{Bib}
\bibliographystyle{ieeetr}

\end{document}